\documentclass[final,5p,times]{elsarticle}

\usepackage[english]{babel}     
\usepackage[T1]{fontenc}        
\usepackage[utf8]{inputenc}     

\usepackage{eurosym}            

\usepackage{amsmath,amsfonts,amssymb,bm} 
\usepackage[per-mode = symbol]{siunitx}
\DeclareSIUnit{\EUR}{\text{\euro}}

\usepackage[activate={true,nocompatibility},final,tracking=true,kerning=true,spacing=true,factor=1100,stretch=10,shrink=10]{microtype}
\microtypecontext{spacing=nonfrench}

\SetExtraKerning[unit=space]
{encoding={*}, family={*}, series={*}, size={footnotesize,small,normalsize}}
{\textendash={400,400}, 
    "28={ ,150}, 
    "29={150, }, 
    \textquotedblleft={ ,150}, 
    \textquotedblright={150, }} 

\SetExtraKerning[unit=space]
{encoding={*}, family={qhv}, series={b}, size={large,Large}}
{1={-200,-200}, 
    \textendash={400,400}}

\usepackage{graphicx}   
\usepackage{caption}
\usepackage{subcaption} 

\usepackage{epstopdf}   

\usepackage{booktabs}   
\usepackage{multirow}   
\usepackage{threeparttable} 

\usepackage{lineno}
\modulolinenumbers[5]

\usepackage[pdftex]{hyperref}       
\hypersetup{colorlinks          = true,         
            allcolors           = black,        
            bookmarksnumbered   = true,         
            bookmarksopen       = true,         
            breaklinks          = true,         
            pdfdisplaydoctitle  = true,         
            pdfpagelayout       = OneColumn,    
            pdfstartview        = FitH,         
            pdfpagemode         = UseNone,      
            pdftitle            = {Integrating Small Satellite Communication in an Autonomous Vehicle Network: A Case for Oceanography},
            pdfauthor           = {Andre Guerra, Maria Costa, Diego Nodar, Sergio Ferreira},
            pdfsubject          = {Article on why and how to integrate satellite communications in an autonomous vehicle network.}
} 

\journal{Acta Astronautica {\emph{\scriptsize\url{https://doi.org/10.1016/j.actaastro.2018.01.022}}}}

\usepackage{xcolor}         

\usepackage[color=blue!20!white,linecolor=red,figcolor=white,textwidth=3cm,colorinlistoftodos,%
]{todonotes}                                                

\begin{document}


\begin{frontmatter} 
    \title{Integrating Small Satellite Communication in an Autonomous Vehicle Network: A Case for Oceanography}
    
    \author[FCUP]{Andr\'{e} G. C. Guerra\corref{cor1}}
    \ead{aguerra@fc.up.pt}
    
    \author[LSTS]{Ant\'{o}nio S\'{e}rgio Ferreira}
    \ead{asbf@lsts.pt}
    
    \author[LSTS]{Maria Costa}
    \ead{mariacosta@lsts.pt}
    
    \author[VIGO]{Diego Nodar-L\'{o}pez}
    \ead{diego.nodar@space.uvigo.es}
    
    \author[VIGO]{Fernando Aguado Agelet}
    \ead{faguado@tsc.uvigo.es}
    
    \cortext[cor1]{Corresponding author}
    
    \address[FCUP]{Departamento de F\'{i}sica e Astronomia, Centro de F\'{i}sica do Porto, Faculdade de Ci\^{e}ncias, Universidade do Porto, Portugal} %
    
    \address[LSTS]{Laborat\'{o}rio de Sistemas e Tecnologia Subaqu\'{a}tica, Faculdade de Engenharia, Universidade do Porto, Portugal}
    
    \address[VIGO]{Escola de Enxe\~{n}ar\'{i}a de Telecomunicaci\'{o}n, Universidade de Vigo, Espa\~{n}a}
    
    \begin{abstract}
    Small satellites and autonomous vehicles have greatly evolved in the last few decades.
    Hundreds of small satellites have been launched with increasing functionalities, in the last few years.
    Likewise, numerous autonomous vehicles have been built, with decreasing costs and form-factor payloads.
    Here we focus on combining these two multifaceted assets in an incremental way, with an ultimate goal of alleviating the logistical expenses in remote oceanographic operations.
    The first goal is to create a highly reliable and constantly available communication link for a network of autonomous vehicles, taking advantage of the small satellite lower cost, with respect to conventional spacecraft, and its higher flexibility.
    We have developed a test platform as a proving ground for this network, by integrating a satellite software defined radio on an unmanned air vehicle, creating a system of systems, and several tests have been run successfully, over land.
    As soon as the satellite is fully operational, we will start to move towards a cooperative network of autonomous vehicles and small satellites, with application in maritime operations, both \textit{in-situ} and remote sensing.
    \end{abstract}
    
    \begin{keyword}
        Small Satellites \sep Autonomous Vehicles \sep Satellite Communication \sep System of Systems \sep Maritime Operations \sep Field Trials
    \end{keyword}

\end{frontmatter}



%
%
%

\section{Introduction}
\label{sec:Introduction}

Oceans are a fundamental piece of the Earth ecosystem, not only because they cover more than 70\% of the Earth surface, but also by playing a central role in the world climate.
Therefore, performing measurements of all its features, with both large temporal and spatial resolution, is an important task for several disciplines.
Besides scientific objectives, there are also economic and security issues, which most coastal countries have been trying to tackle for several years.
This has been mainly achieved through the use of conventional ships or static assets, which currently allow for several \textit{in situ} methods of measuring ocean features.
Albeit their small scales, the measurements performed are very accurate and can even be used to calibrate Earth observation satellites.

In the last few years, autonomous vehicles, either underwater, surface or aerial, increased the capabilities of traditional assets, by surveying larger areas in less time and at a smaller cost.
The use of autonomous systems provides numerous unparalleled advantages to typical, human-centred, approaches to problem solving when dealing with operational scenarios such as surveillance~\cite{Zhou2010}, search and rescue~\cite{Bitton2008}, oceanographic and/or atmospheric studies~\cite{McGuillivary2012,Fortuna2013}.
Advances in autonomous vehicles are in a state of rapid evolution not only due to the lowering of hardware costs, which impact directly the feasibility of more ambitious endeavours, but also due to the increase in quantity and quality of open-source alternatives to regular commercial autopilot and payload solutions.
These variables have set the stage for an ever increasing plethora of applications for autonomous systems not just centred on a single vehicle paradigm but more inline with a multiple system approach.
Consequently, networked vehicles have been a topic of great interest due to the extended capabilities which can be extracted for problem solving purposes~\cite{Eckert2013}.

Multi-vehicle operational setups have been spearheaded by the LSTS (\textit{Laboratório de Sistemas e Tecnologia Subaquática}) from the University of Porto for quite sometime.
Several developments have culminated in a robust software toolchain and in low-cost vehicle solutions, whose discussion will be further expanded in section~\ref{sec:LSTS}.
Moreover, all integration developments are regularly tried and tested in large scale multiple vehicle deployments~\cite{Sousa2016}.
However, these large scale multiple deployments lead to an increased level of difficulty in terms of guaranteed interoperability in teams of heterogeneous vehicles~\cite{Bagdatli2010,Fuchs2013}.

The challenge of implementing network interoperability is of particular interest since unmanned assets can be seen simultaneously middle and end nodes in a multi-hop network.
Furthermore, there is always a limitation in terms of range and bandwidth of communication.
Therefore, there has been an increasing tendency of providing redundancy for the networked system by implementing different simultaneous communication channels in each asset~\cite{Ferreira2017}.
While allowing to overcome some limitations of each individual communication channel, this added redundancy creates, at the same time, a more robust communication backbone for the underlying vehicle network.
In this regard, and specially when dealing with remote and/or spatially large covering operations, a \textit{de facto} communication element of a multi-channel backbone is the satellite link.
Providing a global and reliable means of communication, while empowering local gateways, or replacing them all together if needed, has long caught the attention of groups working in remote locations, hence positioning satellite communication as a paramount player in robust networking in multi-vehicle scenarios.

Nowadays the most well know satellite communication systems are INMARSAT, Iridium, and Argos, for the particular case of maritime applications.
Some of these systems have a large cost not only for the user to have access to them, but also for the provider to set up and maintain them.
Their characteristics, both in terms of platforms used as well as user terminals, are examined in section~\ref{sec:Current}.

In the last two decades, cheaper spacecraft, which use commercial off-the-shelf (COTS) components, have been growing in number, capabilities and possible applications, the so-called small satellites~\cite{Guerra2016}.
These spacecraft, classified as micro ($<100\si{kg}$), nano ($<10\si{kg}$) and pico ($<1\si{kg}$) satellites, have among other advantages an inherent cost reduction, although still following industry standards.
Therefore, a constellation can be, in principle, constructed at a lower price, assuring thus a continuous cover of some sites or even the globe.
Nevertheless, with the known limitations of small satellites, for instance, the limited power generation capability, they cannot be used for all kinds of communication systems.
An example of a small communication satellite system, created initially to support small scientific and humanitarian operations, is the HumSat system (discussed in section~\ref{sec:HumSat}).
This system is already on the second version, where two way communication is one of the most important new features.

Our main goal is thus to have an integrated operational scenario with autonomous vehicles and small satellites linked and working together.
An overview of this scenario can be found in section~\ref{sec:Missions}.

Firstly, this choice would free us from expensive current satellite communication while, at the same time, maintaining the satellite link as an integral part of the communication backbone.
Secondly, by equipping the spacecraft with other payloads, it would expand the satellite's operational portfolio, making it part of the remote observation system, and helping the autonomous vehicles navigate the monitored area, and ultimately to be ``at the right place at the right time''.

The first step to accomplish our goal, is to have the autonomous vehicles and small satellites talking to each other.
To this end, we integrated a HumSat software defined radio (SDR) on an unnamed air vehicle (UAV).
Several bench tests were carried out, to make sure everything was working as intended, together with some field trials.
The work developed by LSTS and the University of Vigo (UVigo), and the obtained results, are described in section~\ref{sec:Integration}.

By using a SDR, and in particular the HumSat SDR, we have a system that can be reconfigured as required.
For example, any vehicle equipped with the SDR can act as a storage node, if for some reason the satellite is not available, forwarding the data when a link to the ground station or the satellite is available.
Therefore, this flexibility allows to create a system of systems.


%
%
%

\section{LSTS Networked Systems}
\label{sec:LSTS}

Currently the LSTS is composed of a multi-disciplinary research team, including faculty members and students, with Electrical, Computer, and Mechanical Engineering and Computer Science backgrounds.
Vehicles in the LSTS fleet are engineered for networked operations, and use modular hardware and software components to facilitate development, maintenance, and operations.
The LSTS open source software toolchain allows the operators at the LSTS control stations to command and control all types of vehicles in an uniform manner.

\subsection{Software toolchain}
\begin{figure}[!htb]
    \centering
    \includegraphics[width=0.75\columnwidth]{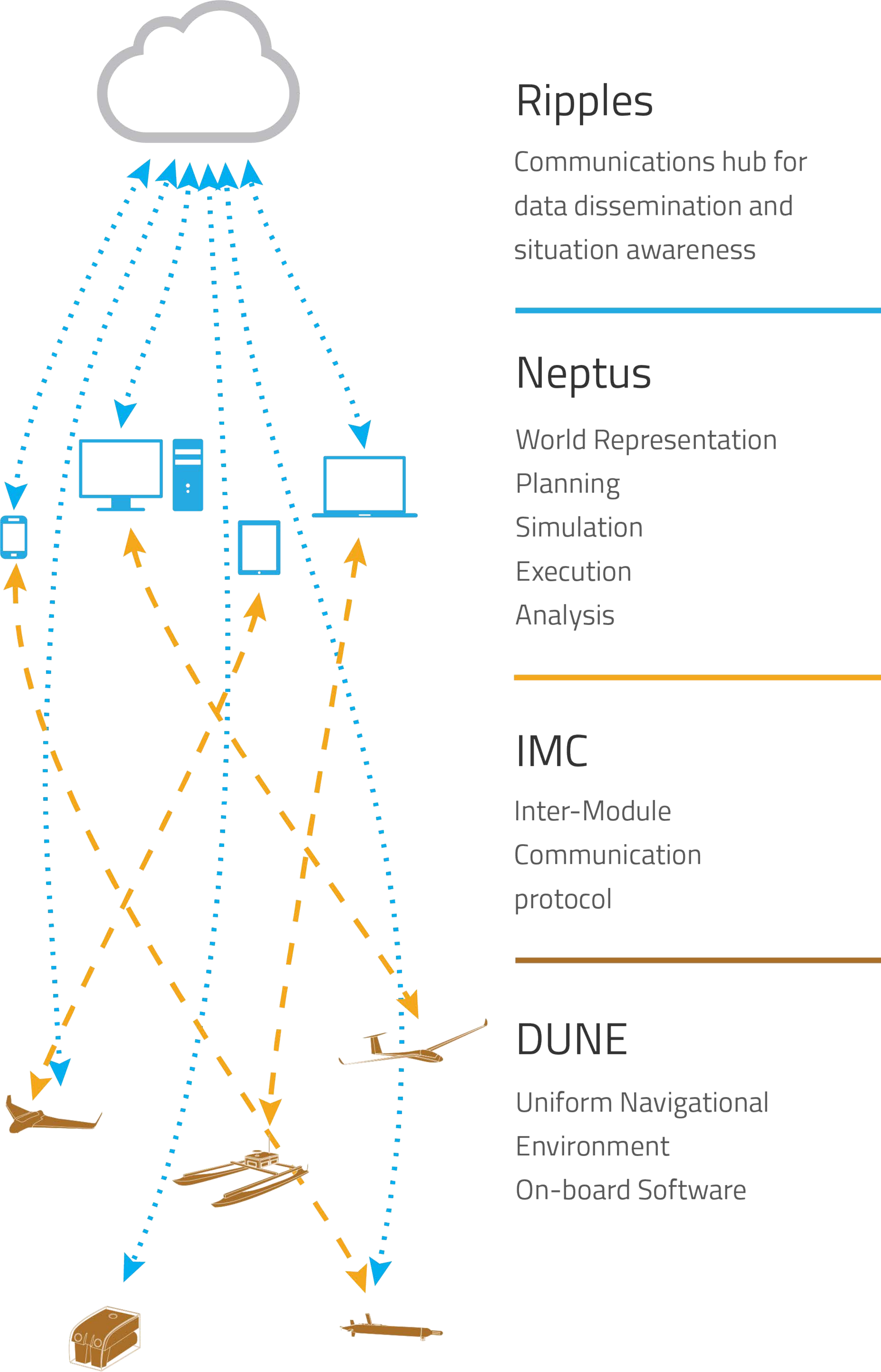}
    \caption{Visual representation of the LSTS toolchain layers.}
    \label{fig:toolchainEN}
\end{figure}

The LSTS toolchain\footnote{The open source software can be found at \url{https://github.com/LSTS}.}, represented in Figure~\ref{fig:toolchainEN}, is an open-source software suite for mixed-initiative control (humans in the planning and control loops)~\cite{Pinto2013} of unmanned ocean and air vehicles operating in communication challenging environments with support for Disruptive Tolerant Networking (DTN) protocols~\cite{Merani2011}.
The unique features of the toolchain are built on experience with the coordinated operation of heterogeneous vehicles~\cite{Ferreira2017}.

\textbf{Ripples} is a communication hub for data dissemination and situation awareness.
It serves as a web entry point for storing situational awareness information like asset positions, mission specifications and collected data.
Ripples is used to collect data from different sources and to maintain a global situation awareness.
All data being sent and received from remote devices is stored in Ripples, enabling coordination among geographically distributed teams, even in communication restricted environments.

\textbf{Neptus} is a distributed command and control infrastructure for the operation of all types of unmanned vehicles.
It supports the different phases of a typical mission life cycle: planning, simulation, execution and post-mission analysis.
It can be adapted by operators to fit mission-specific requirements and extended by developers through a comprehensive plug-in framework.

\textbf{IMC} (Inter-Module Communication) protocol is a message oriented protocol designed and implemented to build interconnected systems of vehicles, sensors and human operators.
These are thus able to pursue common goals cooperatively by exchanging real-time information about the environment and updated objectives.
IMC abstracts hardware and communication heterogeneity by providing a shared set of messages that can be serialized and transferred over different means.

\textbf{DUNE} (Unified Navigational Environment) is an on-board software running on the vehicles.
It is responsible not only for every interaction with sensors, payload and actuators, but also for communication, navigation, control, manoeuvring, plan execution and vehicle supervision.
It is written to be independent of CPU architecture, as well as operating system.

\subsection{Autonomous vehicle system}

Although there are various vehicles available at the LSTS, we opted to start integrating smaller UAVs as our test platform.
This was done due to the simplicity of the hardware integration, and the fact that the communication links on these vehicles are very straightforward in their configurations.

The UAV currently used by LSTS is the X8 Skywalker delta wing.
It is a COTS vehicle, modified at the LSTS, which has been used within various types of operational scenarios in previous projects.

The glider shape and size allows for a logistically light operation setup with a short preparation time (about \SI{20}{minutes} for the first flight, where the UAV is mounted and all consoles turned on, and around \SI{5}{minutes} for subsequent flights).
Furthermore, the vehicle possesses a wide oval internal area, which allows for easy accommodation of different kinds of payload.
This proved important for the physical integration process of the HumSat system, due to its prototype nature.
It is thus a vehicle perfected for low altitude reconnaissance scenarios, with live video feed, and for remote sensor scenarios.

As presented in Table~\ref{tab:X8_Specs}, the main construction material of the X8 is Expanded PolyOlefin (EPO) foam.
The X8 has no documented turbulence restrictions and no minimum visibility conditions for operation.
Moreover, it has no strict outside air temperature limitations, except for the ones we have determined at the LSTS (\SIlist{0;40}{\degreeCelsius}).
Regarding any in-flight icing issues, the LSTS X8 does not operate in any environment that would allow such a scenario.
\begin{table}[!htb]
    \footnotesize
    \centering
    \caption{X8 Vehicle Specifications.}
    \label{tab:X8_Specs}
    \begin{tabular}{cc}
        \toprule
        Parameter                               & Value \\
        \midrule
        Model Designation                       & X8 Skywalker \\
        Design                                  & Fixed wing delta wing \\
        Material                                & EPO Foam \\
        Motor Type                              & Electric \\
        Fuel/Energy                             & LiPo battery \\
        Length                                  & \SI{0.80}{m} \\
        Wingspan                                & \SI{2.12}{m} \\
        Body Diameter                           & Oval (from \SIrange{0.15}{0.30}{m}) \\
        Brakes                                  & None \\
        Hydraulic System                        & None \\
        Parachute                               & None \\
        Weight Empty                            & \SI{1.2}{kg} \\
        Mean Take-Off Weight                    & \SI{4.3}{kg} \\
        Indicated Air Speed (Min|Cruise|Max)    & 15 | 18 | 23 (\si{m/s}) \\
        Wind Tolerance (Head|Cross|Gust)        & 17 | 15 | 19 (\si{m/s}) \\
        Maximum Endurance                       & \SI{50}{min} \\
        Maximum Range                           & \SI{20000}{m} \\
        Maximum Altitude                        & \SI{350}{m} \\
        Maximum rate of climb                   & \SI{2}{m/s} \\
        Maximum rate of descent                 & \SI{3}{m/s} \\
        Maximum bank angle                      & \SI{30}{deg} \\
        Turn radius | Turn rate limits          & \SI{31}{m} | \SI{32}{deg/s} \\
        \bottomrule
    \end{tabular}
\end{table}


%
%
%

\section{Commercial satellite communication systems}
\label{sec:Current}

Most satellite communication systems are designed for large volumes of data, \textit{i.e.} the user has at his disposal video, voice and data communication.
Nevertheless, there are systems designed specifically for performing primarily data collection.
The objective for these systems is to have multiple small ground terminals, transmitting small amounts of information, which are then collected by satellites passing above, and downlinked afterwards to a ground station.

\subsection{INMARSAT}
\label{sec:INMARSAT}
The first dedicated maritime satellite communication service was INMARSAT.
The system is arranged in a four-ocean-region configuration~\cite{Evans1999}, with a total of twelve operational satellites~\cite{Chini2009}.
The new network of spacecraft, called Global Express, is composed of three satellites plus a spare one at a Geostationary orbit (GEO), \textit{i.e.} at an altitude of about \SI{35800}{km}~\cite{Inmarsatplc2013}.
Each satellite has a mass at launch of about \SI{6100}{kg} and measures \SI{6.98 x 33.8}{m}, when deployed.
Five solar panels are used to generate over \SI{13}{kW}, after \SI{15}{years} in orbit, and a Xenon ion propulsion maintains the satellite orbit.
There are several ground stations, distributed over the globe, routing all data received to each customer over the internet.
 
The customer has several different equipments available to use the system, including handheld satellite phones (\textit{e.g.} the IsatPhone 2) or satellite radios (\textit{e.g.} the Fleet One Handset)~\cite{Inmarsatplc2013}.
These radios, attached to fixed antennas weighting \SI{2.5}{kg} and measuring \SI{27.5 x 22.1}{cm}, can provide up to \SI{100}{kbits/s}, consuming around \SI{100}{W}~\cite{Inmarsatplc2014}.
Smaller options, \textit{e.g.} the IsatM2M service, has modems measuring just \SI{12.6 x 12.6 x 4.9}{cm} (for powered vehicles) or \SI{43.2 x 14.7 x 2.5}{cm} (for a modem equipped with batteries)~\cite{Orbcomm2017}.
For the latter, the weights goes up to \SI{1.3}{kg}, including batteries.
Antennas for these modems measure \SI{12.6 x 9.3 x 7.4}{cm}.
Since the satellites are at a high orbit, both terminals need more power to send data.
Therefore, the consumption can go up to \SI{9}{W} when transmitting.

\subsection{Iridium}
\label{sec:Iridium}

One of the most well-known satellite communication services is Iridium. 
The system uses 66 satellites (without counting spares) distributed by 11 orbital planes~\cite{Evans1999}.
Unlike any other communication system, these satellites are cable of changing messages between them, with each satellite linked to four others~\cite{Hobby1998}.

Iridium is build of low Earth orbit (LEO) satellites, at an altitude of approximately \SI{780}{km}.
The new spacecraft, the Iridium NEXT, have life spans closer to \SI{12.5}{years}~\cite{OrbitalATK2015}, and are smaller (\SI{2.4 x 9.4}{m} when deployed), lighter (with a mass of \SI{860}{kg}), and require less power to operate (\SI{2.2}{kW}), when compared to INMARSAT~\cite{OrbitalATK2015}.

The user terminals are commercialised by other companies and are usually small, with omnidirectional antennas and limited power available.
Nowadays, one of the top of the line systems includes a broadband connection and three independent voice lines, using an equipment with \SI{1.35}{kg} and a \SI{12.5}{kg} antenna~\cite{CLSAmerica2016}.
This system claims to have data transfer rates of up to \SI{134}{kbits/s}, consuming up to \SI{31}{W} during data transmission.
There are also small transceivers capable of transmitting short burst data, for instance, for data telemetry (with less than \SI{340}{bytes})~\cite{CLSAmerica2016}.
Data rates for these can reach \SI{2.4}{kbits/s}~\cite{CLSAmerica2016}.
Another commercial example is the RockBlock communication equipment, from Rock Seven, which weights \SI{0.076}{kg}, measures \SI{7.6 x 5.2 x 1.9}{cm}~\cite{RockSeven2016}, and the transceiver consumes less than \SI{1}{W} for transmitting~\cite{CLSAmerica2016}.

The companies that sell the terminals lease the access to the system and charge for each voice or data call (\textit{i.e.} duration or volume of data exchanged).
For example, Rock Seven nowadays charges \SI{10}[\$]{\per{month}} so a transceiver can connect to the system, and about \SI{0.14}[\$]{\per{message}} (each message is composed of a maximum of \SI{50}{bytes})~\cite{RockSeven2014}.
CLS Group charges around \SI{1}{\EUR\per{kbytes}} of data transmitted, for the large bandwidth real-time system~\cite{Yann2017}.
Others charge \SI{0.75}[\$]{\per{minute}} to \SI{1.50}[\$]{\per{minute}}, for outgoing calls, and several dollars per minute for incoming ones.

\subsection{Argos}
\label{sec:Argos}

Argos is operated and managed by the CLS Group, which charges about \SI{4}{\EUR\per{day}}, with a maximum of \SI{63}{\EUR\per{month}}~\cite{Yann2017}.
Version 2 of the system offers global coverage and Doppler location capability, with terminals transmitting messages at a regular interval, without a downlink acknowledgement.
With the new version (Argos 3) two way communication is implemented, and the terminals, now called Platform Messaging Transceivers, only send messages when a satellite is in view~\cite{CLSAmerica2011}.

The lighter Argos 3 equipment available (mass of \SI{0.08}{kg}), is manufactured by Elta (exclusively for CLS), measures \SI[allow-number-unit-breaks=true]{1.8 x 4.3 x 10}{cm} and consumes up to \SI{5}{W}~\cite{Elta2017}.
A heavier option is produced by Kenwood (\SI{2.5 x 6 x 8}{cm} with a mass of \SI{0.16}{kg} and a power consumption of also \SI{5}{W})~\cite{CLSAmerica2009}.
All transmitters use UHF to communicate, and have a structured message composition (with maximum \SI{31}{bytes} of user data per message~\cite{Yann2017}).
Average data rate for both transceivers is about \SI{400}{bits/s}~\cite{Yann2017}.

Argos does not have its own dedicated satellites.
Instead, it depends on other large spacecraft to function (\textit{e.g.} MetOp, \SI{4087}{kg}, and NOAA 5th generation, \SI{2232}{kg})~\cite{EuropeanSpaceAgency2017}.
All of these satellites are sun synchronous (altitude of about \SI{800}{km}) and, although designed for other purposes, they are equipped with Argos payloads, a receiver (\SI{16}{kg}) and a transmitter (\SI{8}{kg}), providing a data rate of up to \SI{4.8}{kbits/s}~\cite{CLSAmerica2011}.
For the orbit of these satellites, and the transmitters used, this means that up to \SI{3.1}{kbytes} are upload each pass (in interactive mode with downlink acknowledgement)~\cite{Yann2017}.
Daily this means an average data transmitted of \SI{10}{kbytes}, depending on the transceiver latitude~\cite{Yann2017}.

The ground segment is composed of three main stations, which receive all user data from the satellites.
Apart from those, there are several regional ones, that perform data relay (\textit{i.e.} receive data if the station and transmitter are under the same footprint).


%
%
%

\section{A different comms approach -- The HumSat system}
\label{sec:HumSat}

\subsection{Introduction}
\label{sec:HumSat_Intro}

Building upon the advantages of small satellites, a new communication system started to be developed in 2011, the HumSat system.
A project supported mainly by the UN office for outer space affairs (UNOOSA) and the European Space Agency (ESA), was initiated as a cooperation between the University of Vigo, the California Polytechnic State University, the National Autonomous University of Mexico, and the Regional Center for Education in Space Science and Technology in Latin America and Caribbean Countries~\cite{Castro2012}.
It is also supported by the Internation Astronautical Federation.

It is a smaller system than all previous ones, still under development, but very similar to the Argos concept.
Table~\ref{tab:satcomm_comparison} shows a comparison of this system with the ones described above.
\begin{table*}[!hbt]
    \scriptsize
    
    \centering
    \begin{threeparttable}
        \setlength\tabcolsep{3.5pt}
        \caption{Comparison between satellite communication services (data from Refs.~\cite{CLSAmerica2016,RockSeven2014,CLSAmerica2011,Yann2017}).}
        \label{tab:satcomm_comparison}
        \begin{tabular}{c c c c c c c}
                \toprule
            System                      
                & Orbit Altitude [km] & Comm. Type & Transmitting Power\tnote{\textdagger}\ \ [W] & Data rate\tnote{\textdagger}\ \ [kbits/s] & Data Amount per Message\tnote{\textdagger}\ \ [bytes] & System of systems \\
            \midrule
            \multirow{2}[0]{*}{INMARSAT} 
                & \multirow{2}[0]{*}{\centering\num{35800} (GEO)}   & Voice & 100   & 100   & --\tnote{$\ast$}         & \multirow{2}[0]{*}{No} \\
                &                                   & Data  & 9     & --\tnote{$\ast$}   & \num{6400} &  \\
            \multirow{2}[0]{*}{Iridium} 
                & \multirow{2}[0]{*}{\centering780 (LEO)}     & Voice & 31    & 134   & -         & \multirow{2}[0]{*}{No} \\
                &                                   & Data  & 1     & 2.4   & 50        &  \\
            Argos 
                & 800 (LEO)                         & Data  & 1     & 4.8   & 31        & No \\
            HumSat 
                & 600 (LEO)                         & Data  & 1     & 1.2   & 32        & Yes \\
            \bottomrule
        \end{tabular}
        \begin{tablenotes}
            \item[\textdagger] Values shown are top limits.
            \item[$\ast$] Information not available.
        \end{tablenotes}
    \end{threeparttable}
\end{table*}

The initial aim was to provide a free non-guaranteed store and forward data system to serve areas not covered by telecommunication services, for humanitarian purposes~\cite{Agelet2015}.
Another objective was to have sensors in remote areas transmitting messages to users, without having to set up a communication network~\cite{Castro2012}.

The system is now moving slowly from a technological demonstrator to commercial applications, the HumSat 2.0~\cite{Agelet2015}.
The plan is not only to maximize the number of users, but to include two-way communication and new added value services, such as data relay between terminals, encrypted communication, prioritization, reliable communications and group of terminals management capabilities.
Different access schemes are being devised, as well as error correction algorithms and fixed and mobile terminals support.
All of them enabled by the deployment of a constellation of small satellites.

All HumSat satellites have been controlled using the same operation software and tools.
UVigo built a ground station at the university, with two redundant stations capable of receiving four simultaneous polarisation signals~\cite{Agelet2015}.
Nevertheless, to command and downlink data from these small satellites it is not necessary a big antenna.
Besides supporting the operation of the two satellites described above, it was also used to communicate with other small satellites.

\subsection{User Segment}
\label{sec:HumSat_User}

The transmitter of the HumSat system is called by UVigo as the user terminal.
There are two options available, either use an automatic or autonomous platform, or to integrate the HumSat transmitter directly on the user equipment.
The first is an almost plug and play solution, which simply relays data coming from the user sensors~\cite{Agelet2015}.
For the second terminal the user needs to prepare both the hardware and software to link the transmitter to its equipment.
We focus here on this last option as it is smaller, lighter and consumes less power.

\begin{figure}[!htb]
    \centering
    \includegraphics[width=0.8\columnwidth]{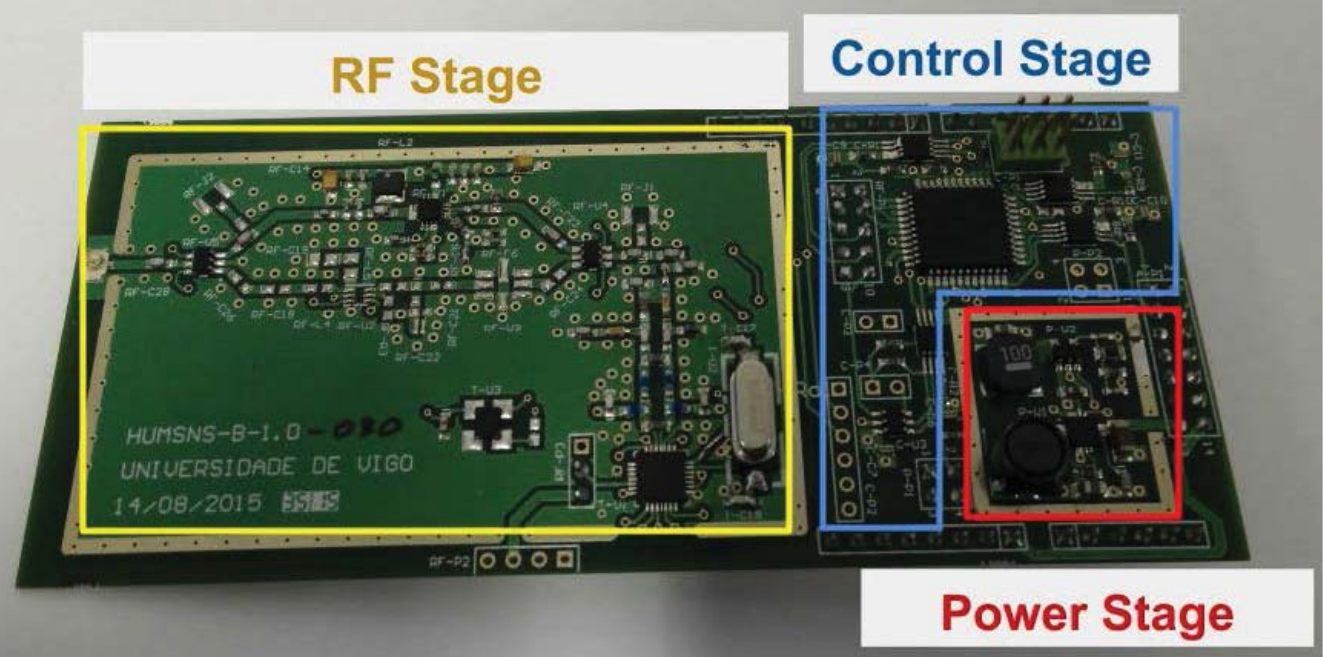}
    \caption{HumSat transmitter photo with labels (from UVigo).}
    \label{subfig:HumSat_UserTerminal}
\end{figure}

This transmitter is a simple board with three modules, a radio frequency (RF) stage, a control stage, and a power stage, as shown in Figure~\ref{subfig:HumSat_UserTerminal}.
The board shown weighs \SI{30}{g} and measures \SI{13 x 5.3}{cm}.
The interface to connect to the board is similar to an Arduino Shield.
The RF connector, to link the terminal to an antenna, is an U.FL connector.

To power the board an unregulated line from \SIlist{4.3;17}{V} has to be used, and the board regulates itself accordingly.
The board consumes in standby less than \SI{0.14}{W}, and up to \SI{3.2}{W} when transmitting at \SI{30}{dBm}.
If just \SI{27}{dBm} are used for transmitting, it consumes \SI{2.4}{W}.

The communication interface is performed over a universal asynchronous receiver/transmitter (UART) serial bus, with a bit rate of \SI{9.6}{kbits/s}.
The RF output signal has a frequency between \SIlist{400;470}{MHz}, and a maximum possible power output of \SI{30}{dBm}, both selected by the user.
The signals have a GMSK (Gaussian Minimum Shift Keying) modulation, with a data rate of up to \SI{1.2}{kbits/s}.

HumSat 2.0 terminals will have a similar electrical configuration.
However, the board is expected to be smaller and capable of higher data rates.
Furthermore, smart terminals are being designed to be dynamically integrated with newer sensors~\cite{Agelet2015}.

\subsection{Space Segment}
\label{sec:HumSat_Space}

Although up to now Argos and HumSat seam very similar, one of the main differences is the spacecraft they use.
HumSat uses a constellation of small satellites, in particular CubeSats, although only a few at a time have been launched yet.

\begin{figure}[!htb]
    \centering
    \includegraphics[width=0.65\columnwidth]{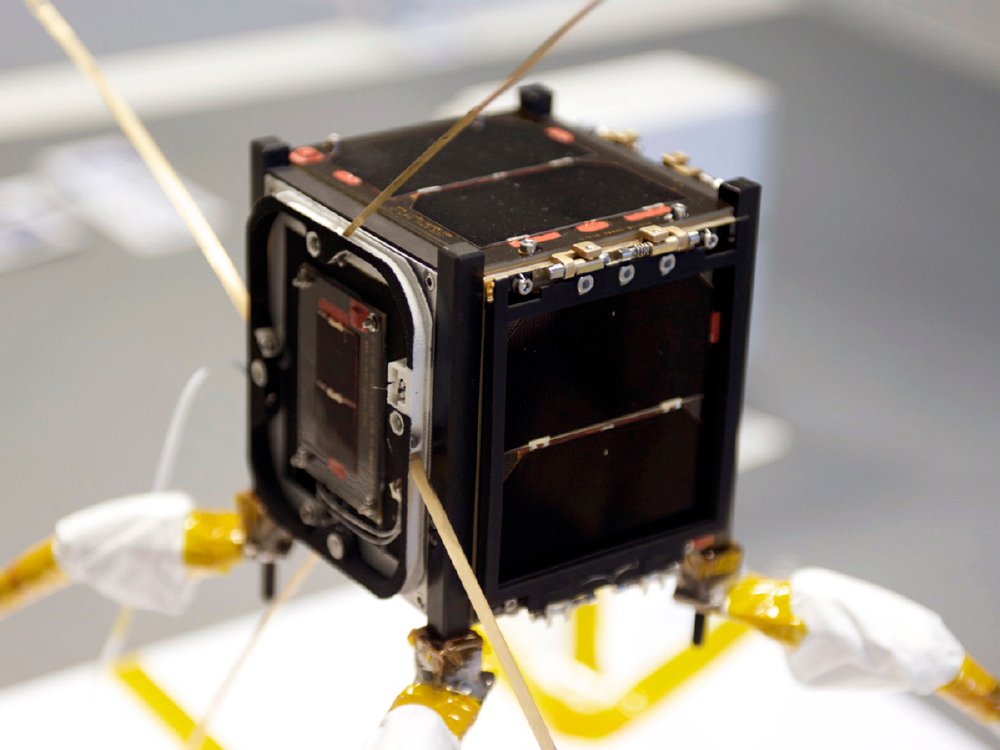}
    \caption{Xatcobeo satellite (from ESA-UVigo/INTA)}
    \label{fig:HumSat_Xatcobeo}
\end{figure}

The first satellite, HumSat-D, was launched in 2013, and was an adaptation of the previous UVigo small satellite, Xatcobeo (Figure~\ref{fig:HumSat_Xatcobeo}).
It was a CubeSat, with \SI{10 x 10 x 10}{cm}, and weighted \SI{1}{kg}.
Using solar cells distributed through most faces, it produced around \SI{3}{W}.
This was sufficient to power the satellite and its communication payload.
This consisted in a single electronics board and one turnstile omnidirectional antenna (with four monopoles).
No attitude control was even necessary to operate the satellite.
Launched from a Dnepr-1 rocket as a secondary payload, its final orbit altitude was \SI{600}{km}, and was operational for over a year~\cite{HumSat_eoPortal}.

With HumSat-D UVigo wanted a demonstrator and guide for the remaining satellites of the system~\cite{Agelet2015}.
The next one was called Serpens, and was built for the University of Brasilia with the support of UVigo, funded by the Brazilian space agency.
This was a bigger CubeSat (\SI{10 x 10 x 30}{cm} and \SI{3}{kg}).
The objective of this satellite was to test different combinations of the payload and user terminals for future implementations.
Additionally, this satellite included an enhanced payload which was able to counteract external interferences in UHF band, detected with the first satellite.


%
%
%

\section{An Integrated System: Mission scenarios}
\label{sec:Missions}

The main driver for the developments presented here is to enable multi-vehicle systems to operate in an integrated and coordinated fashion, by providing them with a robust and stable communication backbone with an ultimate goal of reducing cost and complexity in remote maritime operations.
We endeavour to show that, through the use of new small satellite technology, the underlying cost overhead of using satellite communication can be significantly reduced.
At the same time, we can empower the observation and data collection capabilities of the network, by also having these satellites participate with their own on-board payload.

\begin{figure}[!htb]
    \centering
    \includegraphics[width=1\columnwidth]{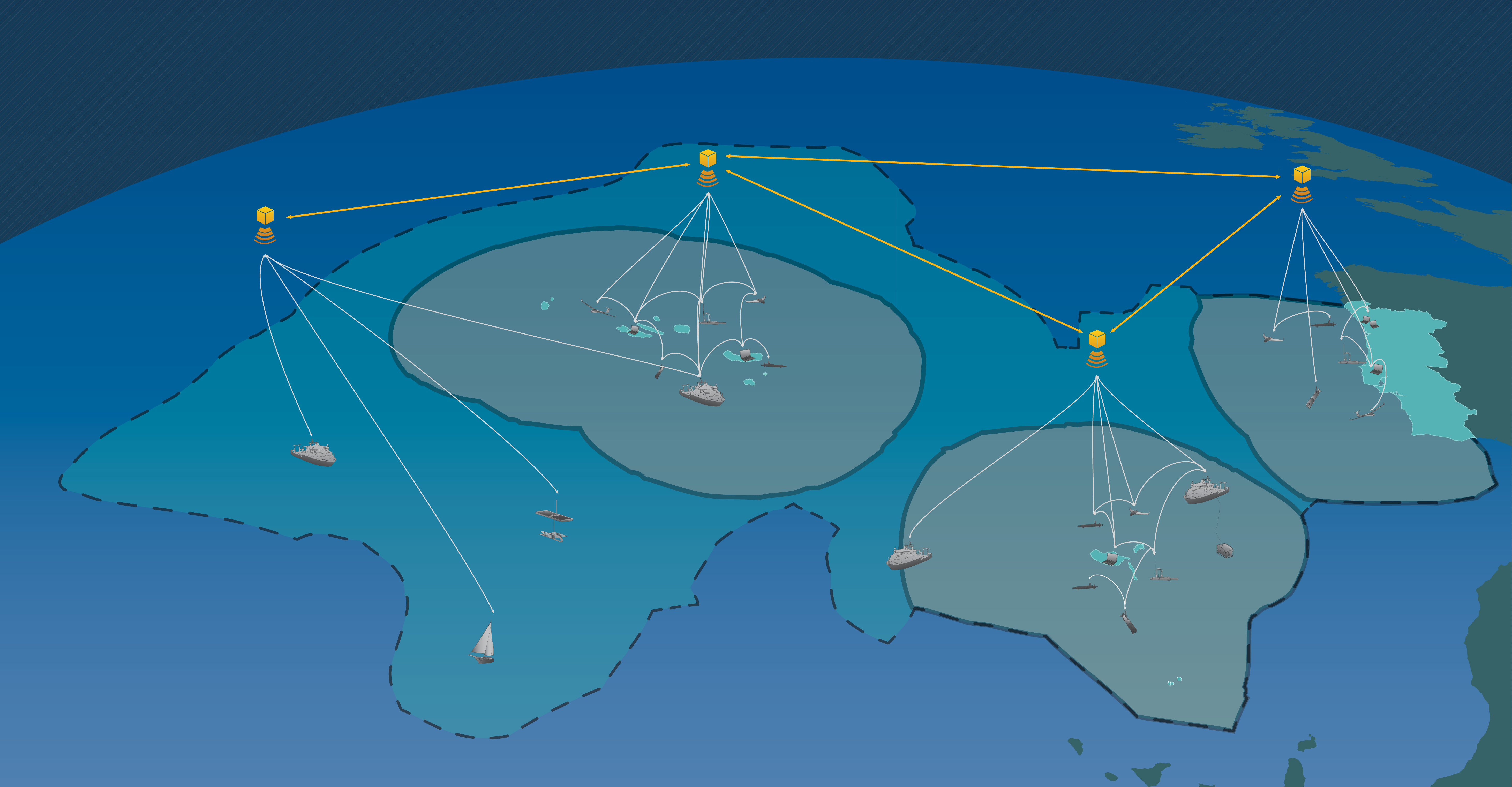}
    \caption{Operational scenario of an integrated network of autonomous vehicles and small satellites.}
    \label{fig:Main_Proposal}
\end{figure}

A pristine example of an operational scenario is the Portuguese continental shelf.
To monitor and carry out surveillance of its current \SI{1.72}{\text{million}\,km^2} area, or the proposed extension of \SI{4}{\text{million}\,km^2} (Figure~\ref{fig:Main_Proposal}), with just conventional methods, represents a large investment, monetary and human.
Therefore, a robust autonomous network like the one described here is fundamental to support these tasks~\cite{Bertolami2017}.

Looking at Figure~\ref{fig:Main_Proposal}, the mission scenario has various layers of active communication links.
These links range from short-range (Wi-Fi/RF), through medium-range (GSM/4G), up to satellite communication.
Moreover, these links are not frozen and can be re-arranged as assets are entering and leaving the operation area.
The plug-and-play modality is fundamental for the viability of a system of systems, and a central reason why the developed efforts here were built upon a pre-existent software tool-chain, devised to handle such operational setups.

Although we plan to start with a single small satellite, a constellation of spacecraft will serve the goal better.
Since this constellation will take time to build, the scalability of the network can be initially probed through the use of various UAVs which, provided with sufficient SDRs, can emulate the satellites.
Although the range and scope of the scenario will not be as far-reaching as with satellites, a lot can be inferred regarding issues of bandwidth and network capacity for the radios.

Even though the project used to fund part of this work is centred in a forest fire project with UVigo~\cite{UniversidadedeVigo2016}, all developments can be easily ported to oceanographic scenarios, since the network created is agnostic to the type of data being streamed and collected.

To facilitate system integration and testing, we began with the LSTS X8 UAV, focusing on transmitting only the vehicle’s position and velocity.
Nevertheless, as the project evolves more metadata will be sent through the satellite, since the communication protocol supports message segmentation and reassembly, to handle information too extensive for a single burst.
The X8 already has Wi-Fi and GSM communication, and we can thus use these for redundancy during the SDR tests.


%
%
%

\section{System integration and test}
\label{sec:Integration}

\subsection{Hardware}
\label{sec:Integration_Hardware}

\begin{figure}[!htb]
    \centering
    \includegraphics[width=1\columnwidth]{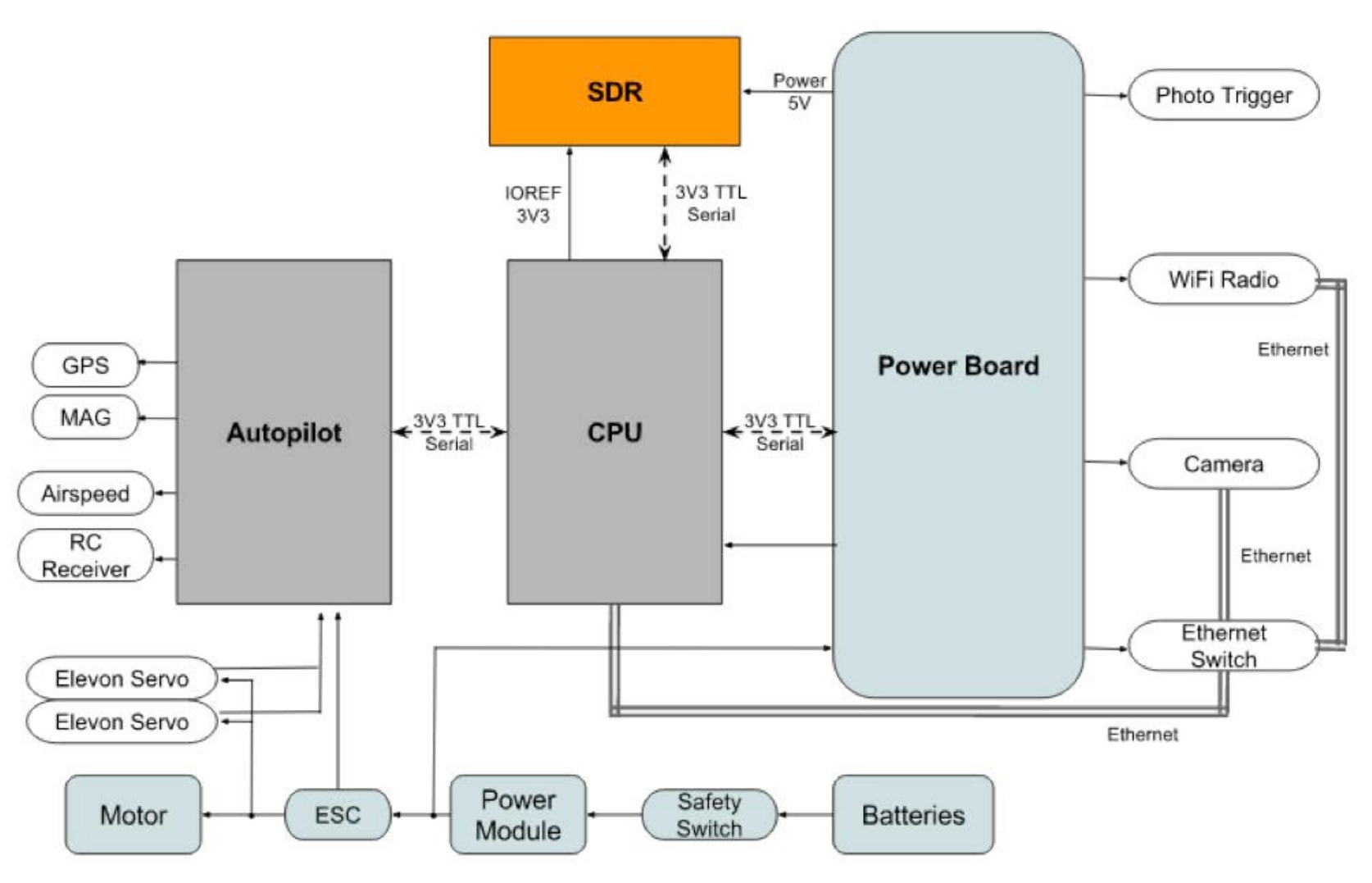}
    \caption{X8 Avionics -- Block Diagram.}
    \label{fig:X8_Avionics}
\end{figure}

Hardware integration of the HumSat system implies adding the SDR board, and an UHF antenna, to the UAV payload.
For the LSTS's X8 delta wing this is quite straightforward, considering both its main CPU and the SDR board have a 3V3 TTL (Transistor–Transistor Logic) serial communication.
Figure~\ref{fig:X8_Avionics} depicts where the SDR is placed within the X8 avionics.

For selecting the antenna, and placing it on the vehicle, the main concern is to match the radiation pattern to our needs.
In particular, the UAV is required to communicate both with a satellite and ground assets, as discussed in section~\ref{sec:Missions}.
Therefore, the selected antenna is a UHF Dipole \SI{433}{MHz}~\cite{FPVMODEL2015}, and is installed on the nose of the vehicle (see Figure~\ref{fig:X8_SDR_Antenna}).

\subsection{Software}
\label{sec:Integration_Software}

All software developments made to integrate the SDR system were done on top of the software toolchain developed at the LSTS, previously described in section~\ref{sec:LSTS}.
A driver task, named SatComms, collects and parses messages, determines when the satellite is in view of the vehicle (and messages can be transmitted), and manages the interaction with the HumSat transmitter (using an adaptation of UVigo's libraries)~\cite{SatCommsProgram2017}.

SatComms falls within the Transports scope (communication and logging) on DUNE's code organization.
During operations the user has full control of the task, being able to change several parameters related to radio configurations, satellite orbital information (period, communication window and last passage), if data is to be transmitted whenever possible, and which type of data.

Bandwidth constraints of the satellite communication system are a major concern, with the current version limited to \SI{32}{bytes} messages.
This is circumvented by fragmenting data into smaller parts and then reassembling them back to its original state.
The two main reasons for this solution are the fact that it was already implemented within the LSTS toolchain, and because it is a message type independent solution.

\subsection{Tests and Results}
\label{sec:Integration_TestingWorkbench}

To test and validate the SDR hardware and software integration, since the real satellite is still to be deployed (2018), UVigo provided a satellite simulator.
This simulator is able to run the same software that will go on-board the future satellite and thus mimics its behaviour.
The test set was divided in three stages: workbench tests, communication dry-run and finally field trials.

\subsubsection{Workbench Tests}

\begin{figure}[!htb]
    \centering
    \includegraphics[width=0.7\columnwidth]{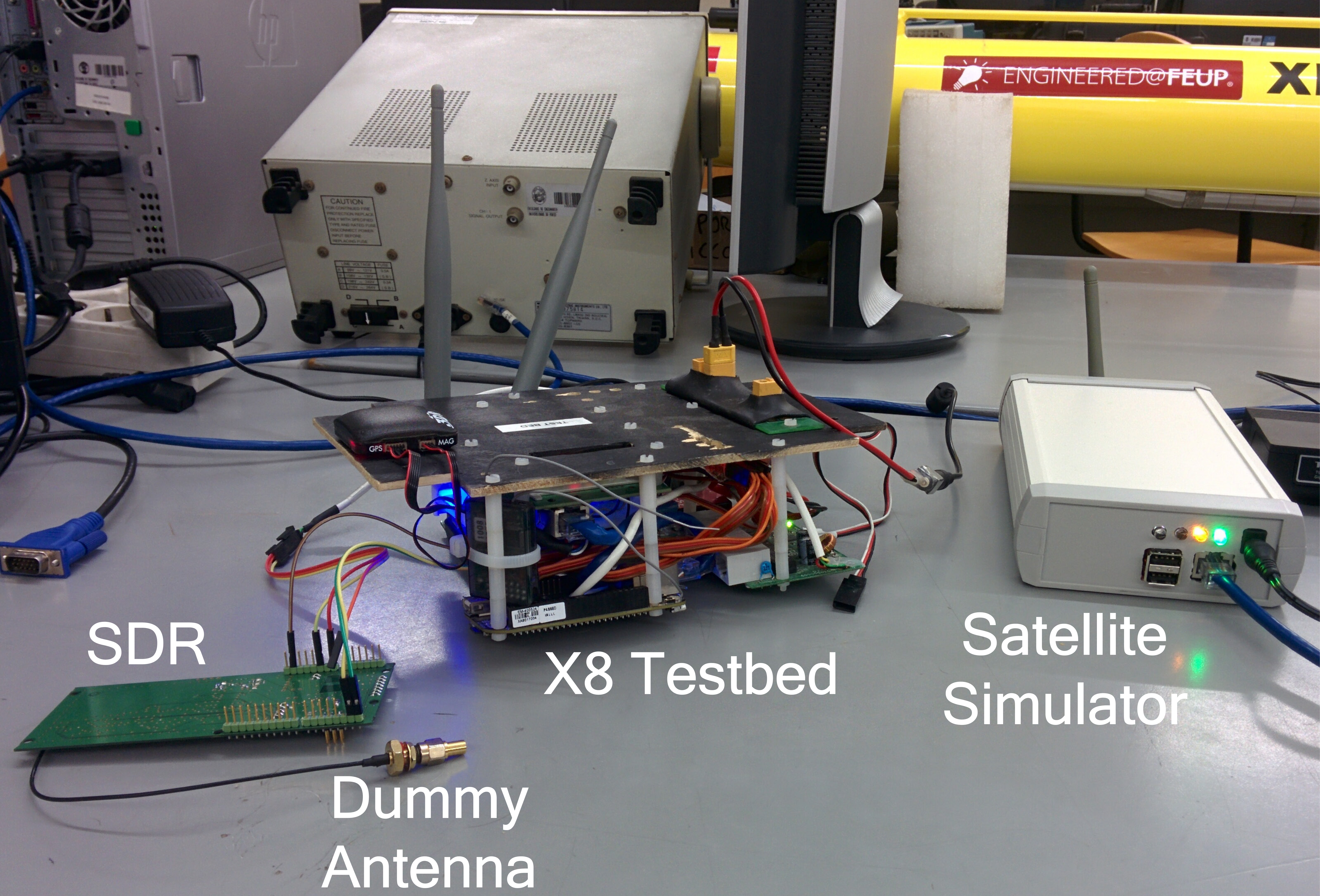}
    \caption{X8 avionics Testbed with HumSat transmitter connected and UVigo satellite simulator.}
    \label{fig:testbed_tests}
\end{figure}

A closed-loop setup was devised to verify every step of the implementation.
It is composed by an exact replica of the UAV avionics (an X8 Testbed), the HumSat SDR transmitter and a dummy antenna, a satellite simulator (provided by UVigo for testing purposes) and a laptop running Neptus (a Neptus Workstation).
Part of this setup is shown in figure~\ref{fig:testbed_tests}.

This Hardware-in-the-Loop (HITL) setup is used to validate the SatComms task behaviour in a simulated environment, in particular the data fragmentation, transmission and reconstruction outcome.
At this stage, all systems are connected in the same network and the simulator is configured to send the incoming data to the Neptus Workstation (IP and Port).
The Neptus Workstation received all message fragments and was able to reconstruct the data.
A few changes were necessary on Neptus to handle repeated message fragments correctly, since the SDR forwards, for redundancy, four times each message.

\subsubsection{Communication Dry-Run}

Following the software developments validation, we setup a test to emulate a real operational scenario.
This requires, apart from the equipment used before, two communication gateways\footnote{The LSTS Manta Communication Gateways is a portable communication hub, allowing multiple communication channels to be active at the same time.}.
Both gateways are capable of providing a 3G connection, each with a different external IP, thus creating isolated networks.
The LSTS server needs to be active as it will serve as a data router.

\begin{figure}[!htb]
    \centering
    \begin{subfigure}[b]{0.5\textwidth}
        \centering
        \includegraphics[width=0.5\textwidth]{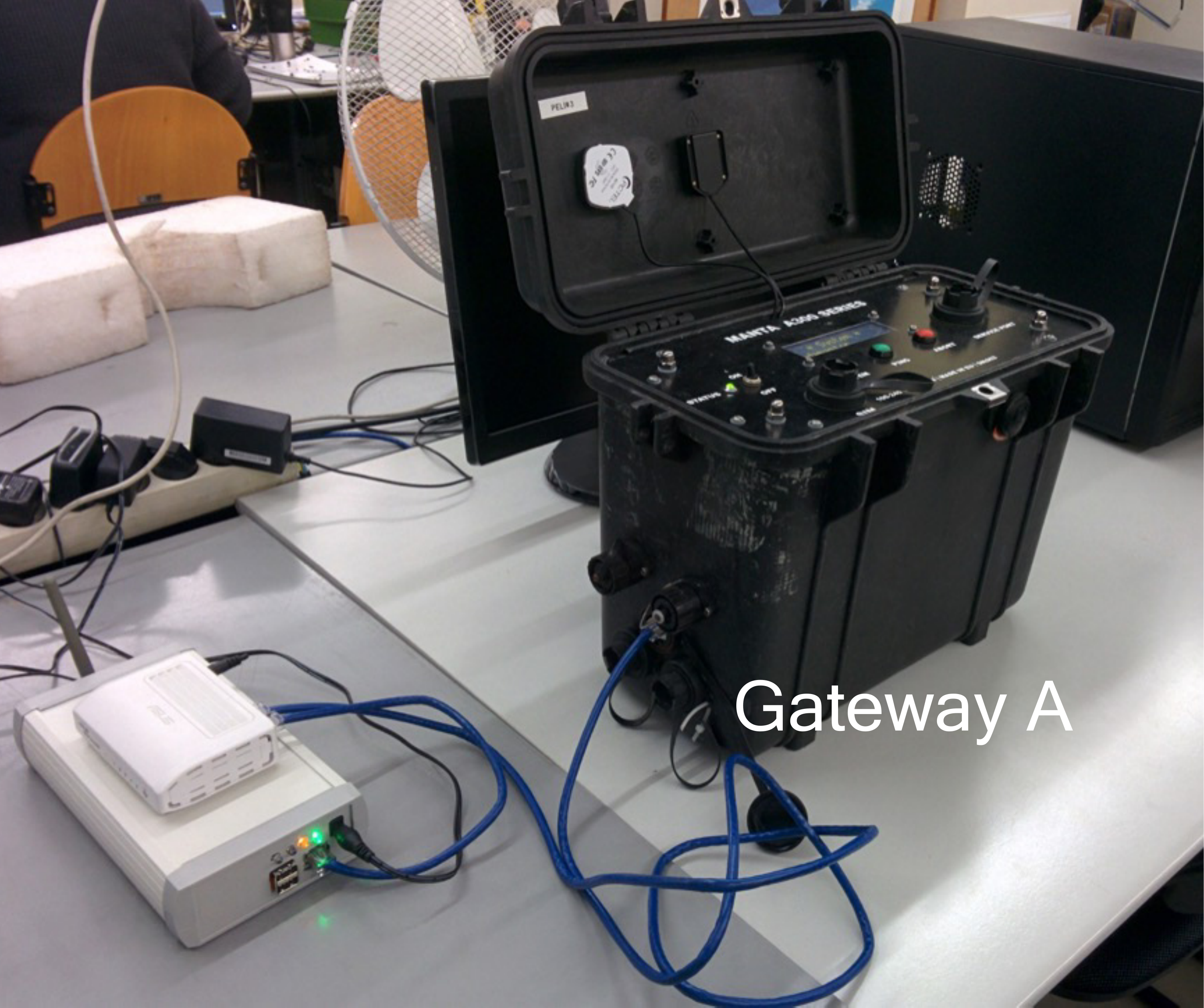}
        \caption{Network A}
        \label{fig:NetworkA_DryRun}
    \end{subfigure}%
    \\
    \begin{subfigure}[b]{0.5\textwidth}
        \centering
        \includegraphics[width=0.8\textwidth]{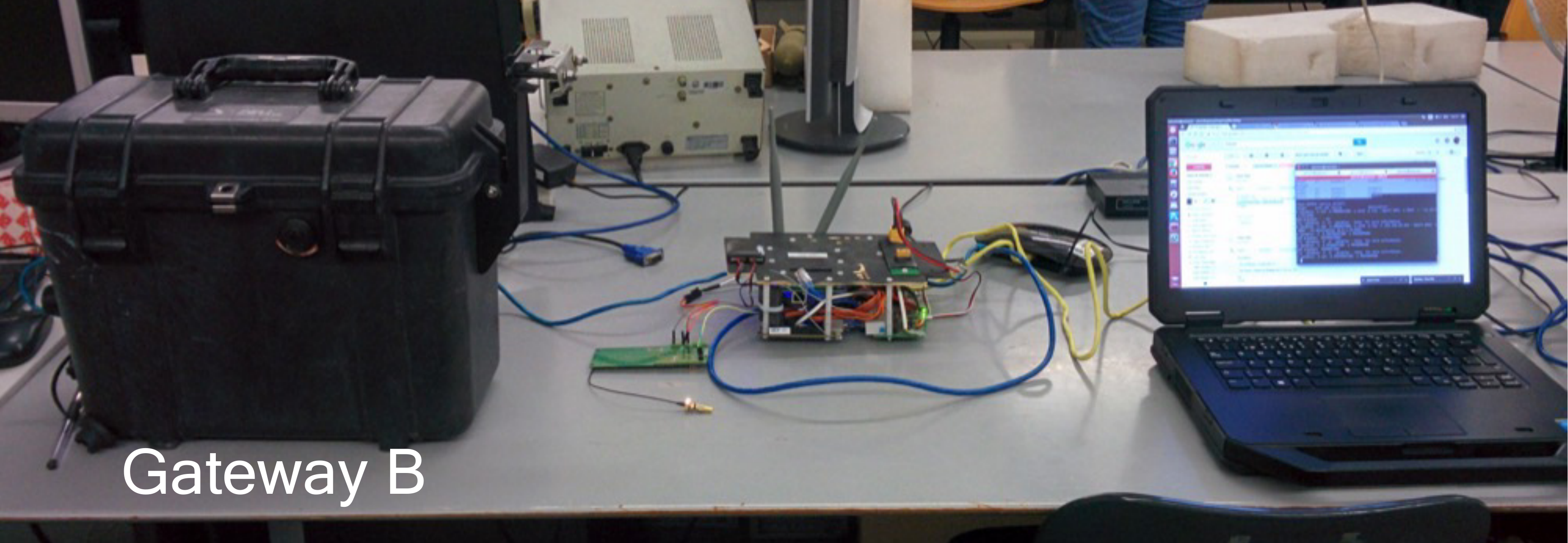}
        \caption{Network B}
        \label{fig:NetworkB_DryRun}
    \end{subfigure}
    \caption{Communication Dry-Run scenario test setup.}
    \label{fig:Comms_Setup}
\end{figure}

In this testing setup, shown in figure~\ref{fig:Comms_Setup}, the simulator is connected to one communication gateway (gateway A), whilst the Testbed (linked to the SDR) and the Neptus Workstation (which will play the part of a remote DPC, Deferred Procedure Call) are connected to another gateway (gateway B).
We placed the Testbed (plus SDR) and the Neptus Workstation on the same network only for logistical convenience, since we also need to have a local workstation to command the Testbed.

\begin{figure}[!htb]
    \centering
    \includegraphics[width=1\columnwidth]{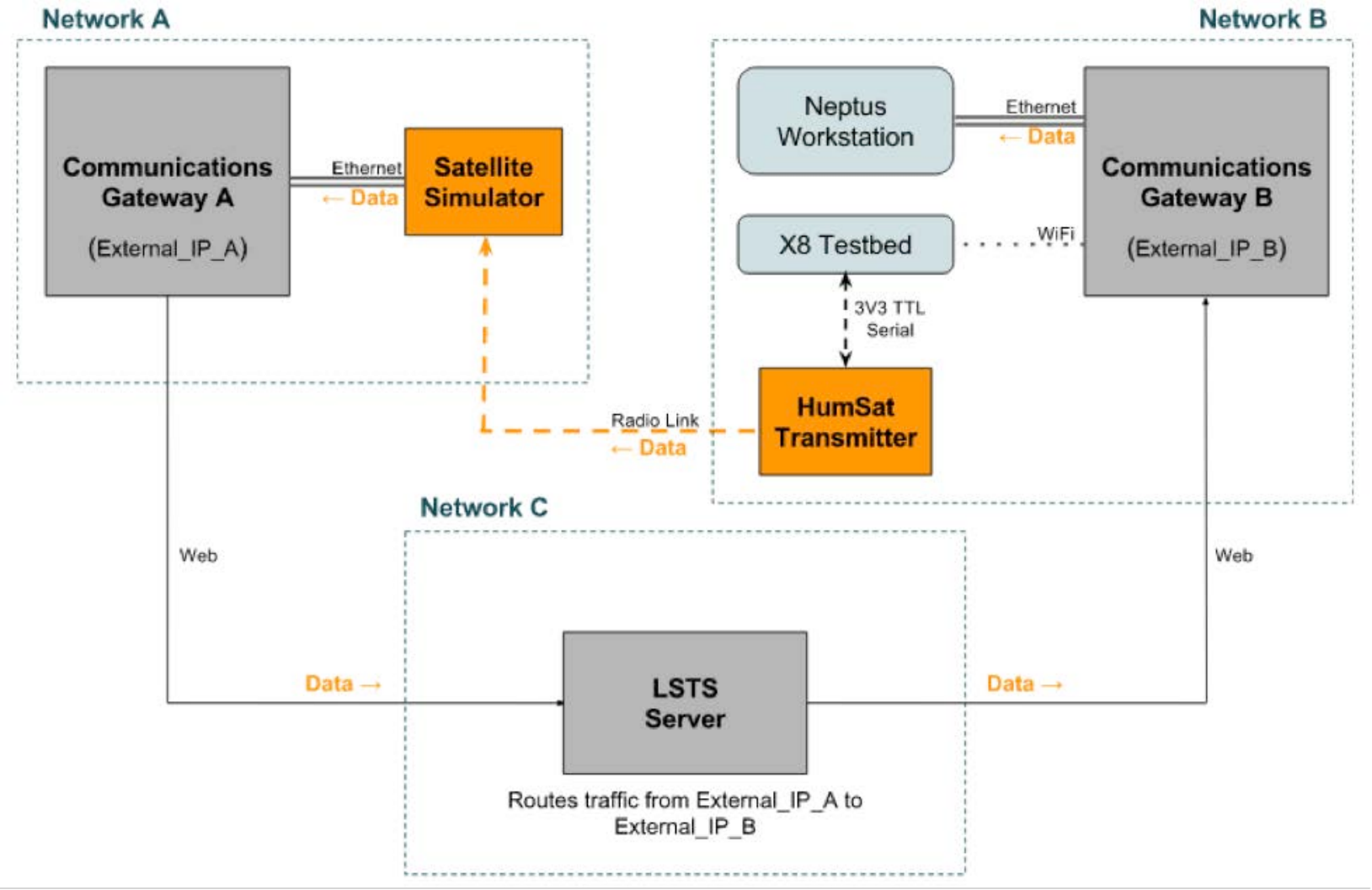}
    \caption{Communication Dry-Run scenario test setup -- Block Diagram.}
    \label{fig:Comms_Diagram}
\end{figure}

The satellite simulator is configured to send the incoming data to the LSTS server (which is in the laboratory connected to the Internet), using gateway A.
In turn, the server routes the data received from the IP of gateway A to the IP of gateway B, which delivers the incoming server traffic to the Neptus Workstation.
A schematic of the links is depicted in figure~\ref{fig:Comms_Diagram}.
The Neptus Workstation can distinguish between the traffic received from the UAV and the one coming from the satellite.
In fact, the only way the Neptus Workstation can receive the data that is being sent with the SDR is through the routing communication done by the LSTS server

A test sequence was carried out to verify everything was working as expected.
The tests made, the desired outcome and the results are enumerated, in sequence, in table~\ref{tab:Comms_DryRun}.
\begin{table*}[!htb]
    \centering
    \begin{threeparttable}[!htb]
        \scriptsize
        \renewcommand{\arraystretch}{1.5}
        \caption{Communication Dry-Run validation tests.}
        \label{tab:Comms_DryRun}
        \begin{tabular}{>{\centering}m{4.5cm} >{\centering}m{5cm} >{\centering}m{6cm} c}
            \toprule
            \textbf{Test}                                        & \textbf{Description }
                &\textbf{Expected Outcome}  
                    & \textbf{Results}\\
            \midrule
            \textbf{Configure network interface in Simulator}    & Manual configuration of eth0 
                & Connection established between gateway and simulator 
                    & OK\\
            \textbf{Configure Simulator \textit{NeptusConnections }}     & Configure IP and port values 
                & Connection established with LSTS server 
                    & OK\\
            \textbf{Connect communication gateways to internet}          & Enable ``Mobile Internet'' in both gateways and give external IPs to POC at LSTS 
                & Mobile Internet successfully established and external IP attributed 
                    & OK\\
            \textbf{Route incoming communication [client]}               & Add iptable rule to route incoming data from LSTS server to Neptus workstations 
                & Iptable rule successfully saved 
                    & OK\\
            \textbf{Route incoming communication [server]}      & Configure LSTS server firewall to route incoming simulator data to field operation network 
                & Connection established between LSTS server and Neptus Workstation 
                    & OK\\
            \textbf{Testbed power up }                           & Power up! 
                & Wait for system in ``SERVICE'' and SatComms task active 
                    & OK\\
            \textbf{Start Simulator}                             & Power up and run script ``LSTS'' 
                & Simulator ready to receive data 
                    & OK\tnote{\textdagger}\\
            \textbf{Configure Satellite}                         & Enable satellite and configure last passage parameter 
                & Set parameter and warns about sent option state 
                    & OK\\
            \textbf{Start transmission!}                         & Enable send option 
                & Wait for satellite to appear and starts sending data 
                    & OK\\
            \textbf{Check transmission}                          & Check if dune is transmitting OK 
                & HumSat terminal confirms transmission OK 
                    & OK\\
            \textbf{Check Simulator reception}                   & Check if simulator is receiving packets 
                & Simulator receiving packets 
                    & OK\\
            \textbf{Check Neptus reception}                      & Check if Neptus is receiving data 
                & Neptus shows incoming MessageParts 
                    & OK\\
            \textbf{Check message reconstruction in Neptus}      & Check if data received is being correctly reconstructed 
                & Neptus shows data reconstructed correctly 
                    & OK\\
            \bottomrule
        \end{tabular}
        \begin{tablenotes}
            \item[\textdagger] Only on second attempt. On the first one, LSTS server was routing data to a port not available, causing the script to crash.
        \end{tablenotes}
    \end{threeparttable}
\end{table*}

\subsubsection{Field Trials}

\begin{figure}[!htb]
    \centering
    \includegraphics[width=0.7\columnwidth]{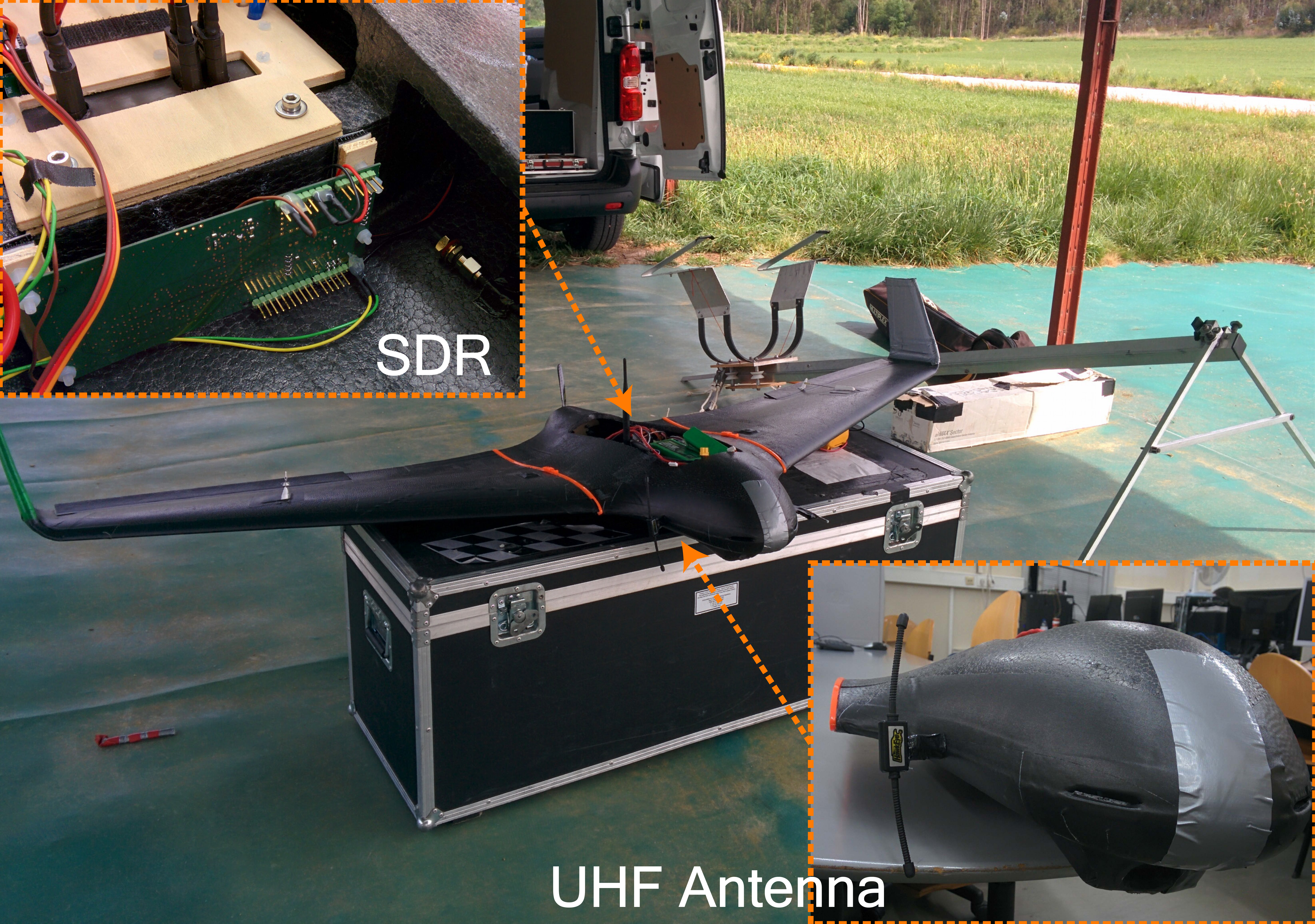}
    \caption{X8 with SDR and UHF antenna installed.}
    \label{fig:X8_SDR_Antenna}
\end{figure}

After testing and evaluating the SDR integration developments in the workbench, the next step was to confirm the results at the airfield.
We installed the HumSat transmitter in the vehicle's belly (vertically, for space optimization purposes) and the UHF antenna on the nose (opposite side to Pitot tube), as shown in figure~\ref{fig:X8_SDR_Antenna}.

\begin{figure}[!htb]
    \centering
    \begin{subfigure}[b]{0.5\textwidth}
        \centering
        \includegraphics[width=0.5\textwidth]{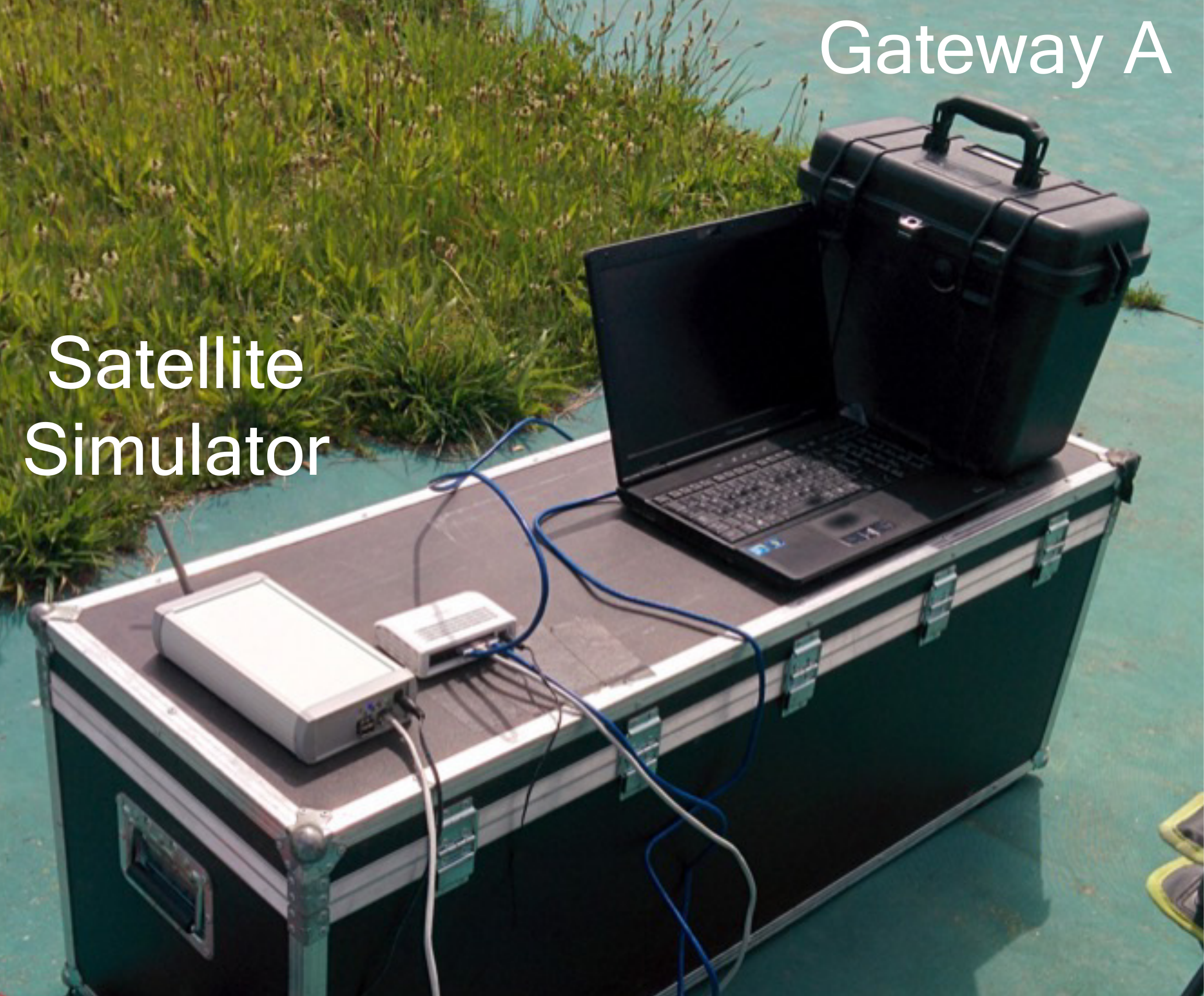}
        \caption{Network A}
        \label{fig:NetworkA_FieldTry}
    \end{subfigure}
    \\
    \begin{subfigure}[b]{0.5\textwidth}
        \centering
        \includegraphics[width=0.5\textwidth]{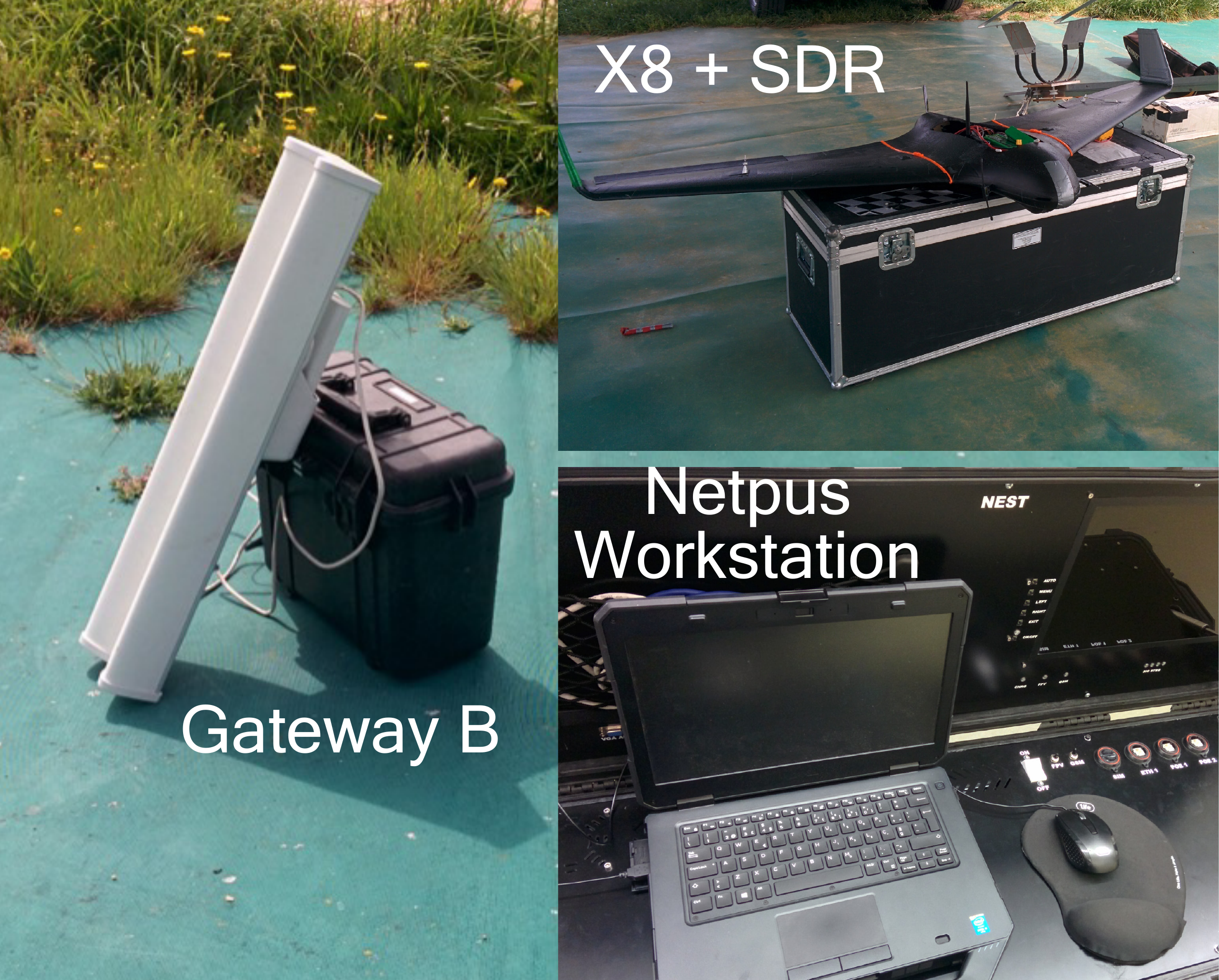}
        \caption{Network B}
        \label{fig:NetworkB_FieldTry}
    \end{subfigure}
    \caption{Field trials operational setup.}
    \label{fig:Trials_Setup}
\end{figure}

The operational setup is identical to the ``Communication Dry-Run'' scenario.
The only difference is the use of the real X8 instead of the X8 Testbed.
Figure~\ref{fig:Trials_Setup} presents the setup used during the field trials, and Table~\ref{tab:Field_Script} displays the scripted field mission step-by-step, with the obtained results during the field trials.

\begin{table*}[!htb]
    \begin{threeparttable}
        \scriptsize
        \renewcommand{\arraystretch}{1.5}
        \centering
        \caption{Field mission script.}
        \label{tab:Field_Script}
        \begin{tabular}{>{}c >{\centering}m{4cm} >{\centering}m{5cm} >{\centering}m{5.5cm} c }
            \toprule
            \textbf{Op. Phase}   & \textbf{Test}  & \textbf{Description} & \textbf{Expected Outcome}  & \textbf{Results}\\
            \midrule
                        & \textbf{Configure network interface in Simulator}
                            & Manual configuration of eth0 
                                & Connection established between gateway and simulator 
                                    & OK\\
            \multirow{5}{*}{\textbf{Setup}}
                        & \textbf{Configure Simulator \textit{NeptusConnections}}
                            & Configure IP and port values 
                                & Connection established with LSTS server 
                                    & OK\\
                        & \textbf{Connect communication gateways to internet}
                            & Enable ``Mobile Internet'' in both gateways and give external IPs to POC at LSTS.
                                & Mobile Internet successfully established and external IP attributed.
                                    & OK\\
                        & \textbf{Route incoming communication [client]}
                            & Add iptable rule to route incoming data from LSTS server to Neptus workstation
                                & Iptable rule successfully saved.
                                    & OK\\
                        & \textbf{Route incoming communication [server]} 
                            & Configure LSTS server firewall to route incoming simulator data to field operation network 
                                & Connection established between LSTS server and Neptus Workstation 
                                	& OK\\
            \cmidrule{2-5}
            \multirow{4}[0]{*}{\textbf{Pre-flight}}
                        & \textbf{Power up UAV} 
                            & Power up UAV and check all systems 
                                & Everything working as expected 
                                    & OK\\
                        & \textbf{Check SDR communication} 
                            & Check if Satcomms is working properly 
                                & Satcomms reports status active 
                                    & OK\\
                        & \textbf{Check SDR range}
                        	& Do SDR + UHF antenna range test on the ground
                            	& Receive data successfully when UAV is away (as far as possible)
                                	& OK\\
                        & \textbf{UAV takeoff} 
                            & Catapult launch 
                                & UAV in air and OK 
                                    & OK\\
            \cmidrule{2-5}
            \multirow{10}[0]{*}{\textbf{Flight}}
                        & \textbf{Survey plan upload}
                            & Make survey plan above desire location and upload it to vehicle 
                                & Vehicle synchronizes plan successfully 
                                    & OK\\
                        & \textbf{Configure Satellite} 
                            & Enable satellite and configure last passage parameter 
                                & Set parameter and warns about sent option state 
                                    & OK\\
                        & \textbf{Start transmission!} 
                            & Enable send option 
                                & Wait for satellite to appear and starts sending data 
                                    & OK\\
                        & \textbf{Survey Plan Execution} 
                            & Start plan previously uploaded 
                                & Plan execution starts
                                    & OK\\
                        & \textbf{Check transmission} 
                            & Check if dune is transmitting OK 
                                & HumSat terminal confirms transmission OK 
                                    & OK\\
                        & \textbf{Check Simulator reception} 
                            & Check if simulator is receiving packets 
                                & Simulator receiving packets 
                                    & OK\\
                        & \textbf{Check Neptus reception} 
                            & Check if Neptus is receiving data 
                                & Neptus shows incoming MessageParts 
                                    & OK\\
                        & \textbf{Check message reconstruction in Neptus} 
                            & Check if data received is being correctly reconstructed 
                                & Neptus shows data reconstructed correctly 
                                    & OK\tnote{\textdagger}\\
                        & \textbf{Waits for end of transmission} 
                            & Loiter away from plan location after survey completed while communication window is open.
                                & Loiter while transmitting.
                                    & OK\\
                        & \textbf{UAV Land} 
                            & UAV retrieval 
                                & UAV landing safely! 
                                    & OK\\
            \bottomrule
        \end{tabular}
        \begin{tablenotes}
            \item[\textdagger] On the first flight there were some problems receiving all MessageParts, since the UAV was forced to fly away from the simulator area due to very strong winds. On the second trial the flight occur at the determined area, and all data was received and reconstructed.
        \end{tablenotes}
    \end{threeparttable}
\end{table*}

During the flight tests we were able to gauge that, with the current HumSat configuration, just one satellite, and a communication window of about \SI{5}{minutes}, approximately \SI{45}{kilobytes} would be sent from the vehicle to the satellite.
In a multi-vehicle scenario, even with a single spacecraft, larger streams of information can be shared among the vehicles, which can then schedule their data transmission, so that a large data packet (\textit{e.g.} an image, an environmental log or some navigational data) can be transmitted and reconstructed on the command and control station.
The field trials were carried out on April 2017, with the next steps already envisioned for mid 2018, when we expect to have a real satellite in operation.


%
%
%

\section{Conclusions}
\label{sec:Conclusions}

In this work we have described the LSTS vehicle system as well as UVigo’s SDR board in detail.
For the LSTS, the toolchain as well as the vehicles have been deployed around the world, with many hours of accumulated operations.
Although the HumSat system from UVigo is more recent, it has been validated with the launch of two small satellites, and several ground terminals deployed in places such as Antarctica, South and North America, and Europe.

Additionally, we characterised some of the current commercial satellite communication systems.
Even though there are several options available, we focus on the ones that could be representative of the different options for sea operational scenarios.
Moreover, we discussed a large bandwidth and real time system, INMARSAT, one with limited bandwidth and gaps in the communication, Argos, and a mixed system that can offer both options and does not require a large transmitting power, Iridium.
These solutions depend on the user's objectives, and have a wide range of prices.

Finally, we have presented the efforts of integrating the HumSat SDR system on-board of a small, low cost, UAV.
The development team created additional software for the LSTS toolchain, which allows the use of the SDR as a viable communication channel, in spite of the current limitations to the amount of data which can be streamed through this channel.
Moreover, a preliminary analysis of the bandwidth issues with the SDR technology was assessed, and important conclusions for the next version of the radio prototype were drawn.
These will influence the next prototype iteration of the SRD system.

\section{Future Work}
\label{sec:Future}

The modifications to the current proposed solution deal with both hardware and software components.
One of the main issues, already being addressed, is the possibility of upgrading the HumSat SDR to allow centralised full-duplex communication, since the current prototype only has one-way.
This would enable each vehicle to receive as well as transmit data, increasing the robustness of the communication network, without using two SDRs.
It would also simplify communication logistics and accommodation problems in small UAVs.
Another issue, is the number of messages that have to be sent to the satellite to transmit a simple variable as the vehicle's estimated state.
Since the main reason, apart from the HumSat \SI{32}{bytes} data length, is the overhead created by the IMC communication protocol, steps are being taken to employ a minimalistic version of this protocol for the satellite channel.
With this approach, less messages would have to be transmitted for each package of data, maximising thus successful variable reconstruction on the Neptus Workstation (\textit{i.e.} the UAV ground station).
All of these solutions will be tested in the next batch of field flight tests.

Although a fully fledged test will be difficult without a complete satellite constellation in place, the scalability of the proposed solution can still be inferred by the use of multiple UAVs in the operation area as smaller relay nodes.
These will help identifying issues about the network before having the logistical costs associated with the spacecraft launches while, at the same time, allow for evaluating the synergies of the SDR technology with the remaining communication links available to the autonomous vehicle teams.

The reality of creating a fully multimedia satellite service would require a considerable initial investment, even taking into account the reduced cost of small satellites when compared with conventional commercial ones.
It would have to take into account not only the costs of developing and building the necessary number of spacecraft, but also to cope with the logistics involved in launching all small satellites, and making sure that they reached their correct positions.
Nonetheless, a small satellite constellation, providing just low speed data connection will, at least in principle, have a smaller cost.
Moreover, the overall success of the small satellites approach, to keep up with the large communication spacecraft, is correlated with specific pieces of technology which are key for future missions.
Examples of these include new and miniaturised high gain antennae, improved software defined radios, high efficient amplifiers, and better orbit and attitude control.



\section*{Acknowledgements}

The work of Andr\'{e} Guerra is supported by the Funda{\c c}\~ao para a Ci\^encia e a Tecnologia (Portuguese Agency for Research) fellowship PD/BD/113536/2015.
Andr\'{e} Guerra would also like to thank his PhD supervisor, Orfeu Bertolami, for several discussions and suggestions on the subject of this paper, and for reviewing it.
The work of Maria Costa and S\'{e}rgio Ferreira was partially supported by Marinfo (NORTE-01-0145-FEDER-000031) and STRIDE (NORTE-01-0145-FEDER-000033).
The work of Diego Nodar and Fernando Aguado was partially supported by the Spanish Ministry of Economy, Industry and Competitiveness under projects HumSat 2.0 (ESP2013-47935-C4-1-R) and IMONss (ESP2016-79184-R).
The work of Maria Costa, S\'{e}rgio Ferreira, Diego Nodar and Fernando Aguado was partially supported by Interreg Sudoe programme under grant number SOE1/P4/E0437 (Fire-RS project).
The authors would like to thank all the support from the people in the Laborat\'{o}rio de Sistemas e Tecnologia Subaqu\'{a}tica (LSTS) and Centro Gallego de Innovaci\'on Aerospacial (CINAE).

\bibliography{SatComms_BIB}

\end{document}